\documentclass[a4paper]{ifacconf}

\usepackage{graphicx,amsmath,url}      % include this line if your document contains figures
\usepackage[round]{natbib}             % required for bibliography
\def\portugues{1} 
\ifx\portugues\undefined
\def\portugues{0}
\fi

\if\portugues0
   \usepackage[english]{babel}
  \else
   \usepackage[spanish,brazil,english]{babel}
\fi

\usepackage{icomma}
\usepackage{color}
\usepackage{xcolor}

\usepackage[T1]{fontenc}

\usepackage[utf8]{inputenc}

\usepackage{ae}

\if\portugues1
 { }
 { }
 { }
 
\fi

\begin{document}

\if\portugues1

% =====================================================================
% =====================================================================
% USE THIS PART IF THE TEXT IS IN PORTUGUES OR SPANISH
% =====================================================================
% If the manuscript is in Spanish, please change the texts adequately.
% =====================================================================
% 
\selectlanguage{brazil}
	
\begin{frontmatter}

\title{Reprodutibilidade na Simulação de um Circuito Caótico} 

\thanks[footnoteinfo]{Os autores agradecem ao apoio financeiro da Fapemig, CNPq e Capes.}

\author[First]{Thalita E. Nazaré} 
\author[Second]{Erivelton G. Nepomuceno}

\address[First]{GCOM - Grupo de Controle e Modelagem \\ UFSJ - Universidade Federal de São João del-Rei\\ Pça. Frei Orlando, 170 - Centro - 36307-352 - São João del-Rei, MG, Brasil (e-mail: thalitanazare@gmail.com).}
\address[Second]{GCOM - Grupo de Controle e Modelagem \\ UFSJ - Universidade Federal de São João del-Rei\\ Pça. Frei Orlando, 170 - Centro - 36307-352 - São João del-Rei, MG, Brasil (e-mail: nepomuceno@ufsj.edu.br)}

\selectlanguage{english}
\renewcommand{\abstractname}{{\bf Abstract:~}}
\begin{abstract}                % Abstract of not more than 250 words.
An evergreen scientific feature is the ability for scientific works to be reproduced. This feature allows researchers to understand, enhance, or even question works that have been developed by other scientists. In control theory the importance of modeling and simulation of systems is widely recognized. Despite this recognition, less attention is paid to the effects of finite precision of computers on the simulation reproducibility of nonlinear dynamic systems. In this work, a case study of reproducibility is presented in the simulation of a chaotic Jerk circuit, using the software LtSpice. In order to do so, we performed simulations of the circuit in the same version of the software on different computers, in order to collect the data and compare them with experimental results. The comparison was made with the NRMSE (Normalized Root Mean Square Error), in order to identify the computer with the highest prediction horizon. Tests performed in 4 different configurations showed the difficulties of simulation reproducibility in LtSpice. The methodology developed was efficient in identifying the computer with better performance, which allows applying it to other cases in the literature.
\vskip 1mm% não altere esse espaçamento
\selectlanguage{brazil}
{\noindent \bf Resumo}:  Um dos aspectos essenciais da ciência é a habilidade de que trabalhos científicos possam ser reproduzidos. Essa característica permite que pesquisadores possam compreender, aprimorar ou mesmo questionar trabalhos que foram desenvolvidos por outros cientistas. Na teoria de controle é amplamente reconhecido a importância da modelagem e simulação de sistemas. Apesar deste reconhecimento, percebe-se uma menor atenção nos efeitos da precisão finita dos computadores na reprodutibilidade de simulação de sistemas dinâmicos não-lineares. Neste trabalho, apresenta-se um estudo de caso da reprodutibilidade na simulação de um circuito caótico Jerk, utilizando o software LtSpice. Para isso foram realizadas simulações do circuito em uma mesma versão do software em diferentes computadores, a fim de coletar os dados e compará-los com resultados experimentais. A comparação foi feita com o índice NRMSE (\textit{Normalized Root Mean Square Error}), com o objetivo de identificar o computador com o maior horizonte de predição. Testes realizados em 4 configurações distintas evidenciaram as dificuldades da reprodutibilidade de simulação no LtSpice. A metodologia desenvolvida foi eficiente em identificar o computador com melhor desempenho, o que permite aplicá-la para outros casos da literatura.
\end{abstract}

\selectlanguage{english}

\begin{keyword}
Reproducibility; Chaotic Circuit; Chaos; Jerk Dynamics; LtSpice. 

\vskip 1mm% não altere esse espaçamento
\selectlanguage{brazil}
{\noindent\it Palavras-chaves:} Reprodutibilidade; Circuitos Caóticos; Caos; Dinâmica Jerk; LtSpice.
\end{keyword}

\selectlanguage{brazil}

\end{frontmatter}
\else

\begin{frontmatter}

\title{Style for SBA Conferences \& Symposia: Use Title Case for
  Paper Title\thanksref{footnoteinfo}} 

\thanks[footnoteinfo]{Sponsor and financial support acknowledgment
goes here. Paper titles should be written in uppercase and lowercase
letters, not all uppercase.}

\author[First]{First A. Author} 
\author[Second]{Second B. Author, Jr.} 
\author[Third]{Third C. Author}

\address[First]{Faculdade de Engenharia Elétrica, Universidade do Triângulo, MG, (e-mail: autor1@faceg@univt.br).}
\address[Second]{Faculdade de Engenharia de Controle \& Automação, Universidade do Futuro, RJ (e-mail: autor2@feca.unifutu.rj)}
\address[Third]{Electrical Engineering Department, 
   Seoul National University, Seoul, Korea, (e-mail: author3@snu.ac.kr)}
   
\renewcommand{\abstractname}{{\bf Abstract:~}}   
   
\begin{abstract}                % Abstract of not more than 250 words.
These instructions give you guidelines for preparing papers for IFAC
technical meetings. Please use this document as a template to prepare
your manuscript. For submission guidelines, follow instructions on
paper submission system as well as the event website.
\end{abstract}

\begin{keyword}
Five to ten keywords, preferably chosen from the IFAC keyword list.
\end{keyword}

\end{frontmatter}
\fi

\section{Introdução}

Um dos aspectos essenciais da ciência é a capacidade de reproduzir resultados científicos obtidos por outros cientistas. Esta perspectiva tem sido tema de discussão na área de teoria de controle e automação em diversas vertentes. Por exemplo, \cite{Bonsignorio2017} afirma que a reprodutibilidade de resultados experimentais é uma característica chave do método científico, mas que apesar disso, a área de robótica e inteligência artificial não tem priorizado essa perspectiva. Nesta mesma linha, o painel de editores do IEEE conduzido pelo editor chefe da  \textit{IEEE Robotics and Automation Magazine} afirmou que a reprodutibilidade na pesquisa científica é imperativa, mas que infelizmente há diversos trabalhos científicos que a reprodução dos resultados é difícil ou mesmo impossível \cite{IEEE2018}.

Na área de teoria de controle é amplamente reconhecido a relevância da modelagem e simulação de sistemas \cite{MNB2013}. Apesar deste reconhecimento, percebe-se uma menor atenção nos efeitos da precisão finita dos computadores na reprodutibilidade de simulação de sistemas dinâmicos não-lineares \cite{Nep2014}. De fato, experimentos numéricos têm sido realizados desde os trabalhos de \citep{lorenz} com a intenção de compreender o comportamento de sistemas dinâmicos não-lineares \citep{Pec2004,HYG1987}.

Em muitos estudos já se pode encontrar a importância da comparação entre análise física e computacional, como vários trabalhos que surgiram após o desenvolvimento do circuito de Chua \citep{chua1992}. Recentemente, esses estudos têm sido motivados pela descoberta de equações diferenciais ordinárias de terceira ordem, cujas soluções apresentam comportamento caótico.  No trabalho \textit{A new class of chaotic circuit} \citep{Spr2000}, Sprott  estabelece uma nova classe de circuitos eletrônicos caóticos composto por resistores, capacitores, diodos e amplificadores operacionais, em que ele apresenta esta comparação entre resultados experimentais e computacionais. 

Desta forma, é possível afirmar que a computação numérica é de fundamental importância na análise de circuitos eletrônicos que descrevem sistemas dinâmicos não-lineares \citep{Mat1987,Spr2011}. Essas análises fazem uso de softwares populares e computadores de fácil acesso a maioria dos pesquisadores. Porém, há muitos trabalhos publicados cuja confiabilidade dos resultados numéricos não tem sido avaliada com cuidado \citep{Loz2013}. 

Entre os trabalhos encontrados na literatura que questionam a confiabilidade das simulações numéricas, em especial as que envolvem sistemas dinâmicos caóticos, encontram-se: \cite{Heitor}, em que os autores apresentam a computação por intervalos como uma forma de se obter resultados mais confiáveis na simulação do mapa logístico. \cite{Paiva2015} e \cite{Naz2017} apresentam a influência da condição inicial e a imprecisão numérica no resultado da simulação do mapa logístico e quadrático, respectivamente. Recentemente, \cite{NM2016} demonstraram o teorema do limite inferior do erro \citep{NMAR2017} para estimar erro em funções recursivas, como os modelos polinomiais NARMAX. Esse teorema mostrou-se útil para testar a confiabilidade de simulações numéricas para sistemas dinâmicos não-lineares. Outros exemplos que utilizam computação intervalar como forma de obter melhores resultados são: \cite{SEP,Marcia}.

Entretanto, trabalhos que utilizam \textit{softwares} de simulação do tipo Spice para análise de circuitos caóticos não apresentam, em sua maioria, a versão do \textit{software} e as configurações do computador onde foram realizadas as simulações.  Esse fato dificulta futuras análises do mesmo circuito, caso esse não apresente uma boa reprodutibilidade. Como exemplo, tem-se o trabalho realizado por \cite{Salamon}, em que os autores identificaram problemas ao reproduzir simulações do Circuito de Chua em diferentes computadores utilizando o \textit{software} Multsim. Há também o trabalho de \cite{Samir} que apresenta a influência de \textit{softwares} e sistemas operacionais em simulações de modelos dinâmicos não-lineares.

Motivando-se no problema de repetibilidade no circuito de Chua esse projeto teve por objetivo verificar se o mesmo ocorre com o circuito caótico Jerk \citep{Spr2011}, comparando dados obtidos em diferentes computadores utilizando o \textit{software} LtSpice e coletados experimentalmente. Esta comparação foi realizada utilizando o cálculo do índice NRMSE, aquele computador que apresentou o menor valor de índice foi escolhido como referência para a análise da reprodutibilidade do LtSpice. Objetiva-se com esse índice identificar o computador que apresente o maior horizonte de predição. Também verifica-se assim, a dificuldade de reprodução de resultados ao utilizar computadores com diferentes configurações.

O restante do trabalho está organizado da seguinte forma. Na Seção 2 são apresentados conceitos básicos do artigo. A Seção 3 apresenta a metodologia abordada no trabalho. Em seguida, na Seção 4, os resultados obtidos são descritos e analisados. Por fim, a Seção 5 apresenta a conclusão e perspectivas de trabalhos futuros.

\section{Conceitos Preliminares}

\subsection{Dinâmica Jerk}
Entre os circuitos que apresentam comportamento caótico, é possível encontrar modelos que foram motivados por equações diferenciais de terceira ordem, representadas pela Eq.(\ref{eqjerk}). A solução para esta equação pode ser encontrada em \citep{Sprott1997}. 
A função não-linear \textit{J} é chamada de ``Jerk'', porque descreve a derivada de terceira ordem de \textit{x}, que corresponde à primeira derivada da aceleração em um sistema mecânico \citep{Spr2011}.
\begin{equation}
\label{eqjerk}
    \dddot{x}=J(\ddot{x},\dot{x},x)
\end{equation}
 A função \textit{J} pode ser representada da seguinte forma, 
 \begin{equation}
     J=-A\ddot{x}-x\mp \dot{x}^2
 \end{equation}
 que é a equação diferencial ordinária mais simples com uma não-linearidade quadrática que apresenta resultado caótico. O termo não-linear de \textit{J} é $\dot{x}$. Nesta equação \textit{A} é o parâmetro de bifurcação que apresenta caos entre a maior faixa dos limites de $2,0168...< A < 2,0577...$ \cite{Spr2011}.

\subsection{Raíz do Erro Quadrático Normalizado}
NRMSE (\textit{Normalized Root Mean Square Error}), pode ser calculado da seguinte forma:
$$NRMSE=\frac{\sqrt{\sum_{k=1}^{N}\left [ y(k)-\widehat{y}(k) \right ]^2}}{\sqrt{\sum_{k=1}^{N}\left [ y(k)-\overline{y} \right ]^2}},$$
em que $y$ são os dados coletados,  $\widehat{y}$ dados simulados, $\overline{y}$ é a média dos dados simulados e $N$ a quantidade de pontos analisados.

\section{Metodologia}

Como mencionado esse trabalho teve por objetivo reproduzir os resultados apresentados por \cite{Spr2011} simulando e implementando o circuito eletrônico proposto em seu artigo. Esta simulação foi realizada em diferentes sistemas operacionais e processadores, permitindo então analisar os resultados obtidos pela implementação prática e em cada computador, verificando assim a reprodutibilidade do LtSpice. Entende-se reprodutibilidade como a capacidade de reprodução dos mesmos resultados de simulação quando sujeitos as mesmas condições iniciais, parâmetros do circuito e parâmetros de configuração do método numérico. 

\begin{figure}[htp]
	\includegraphics[width=.45\textwidth,height=.28\textheight]{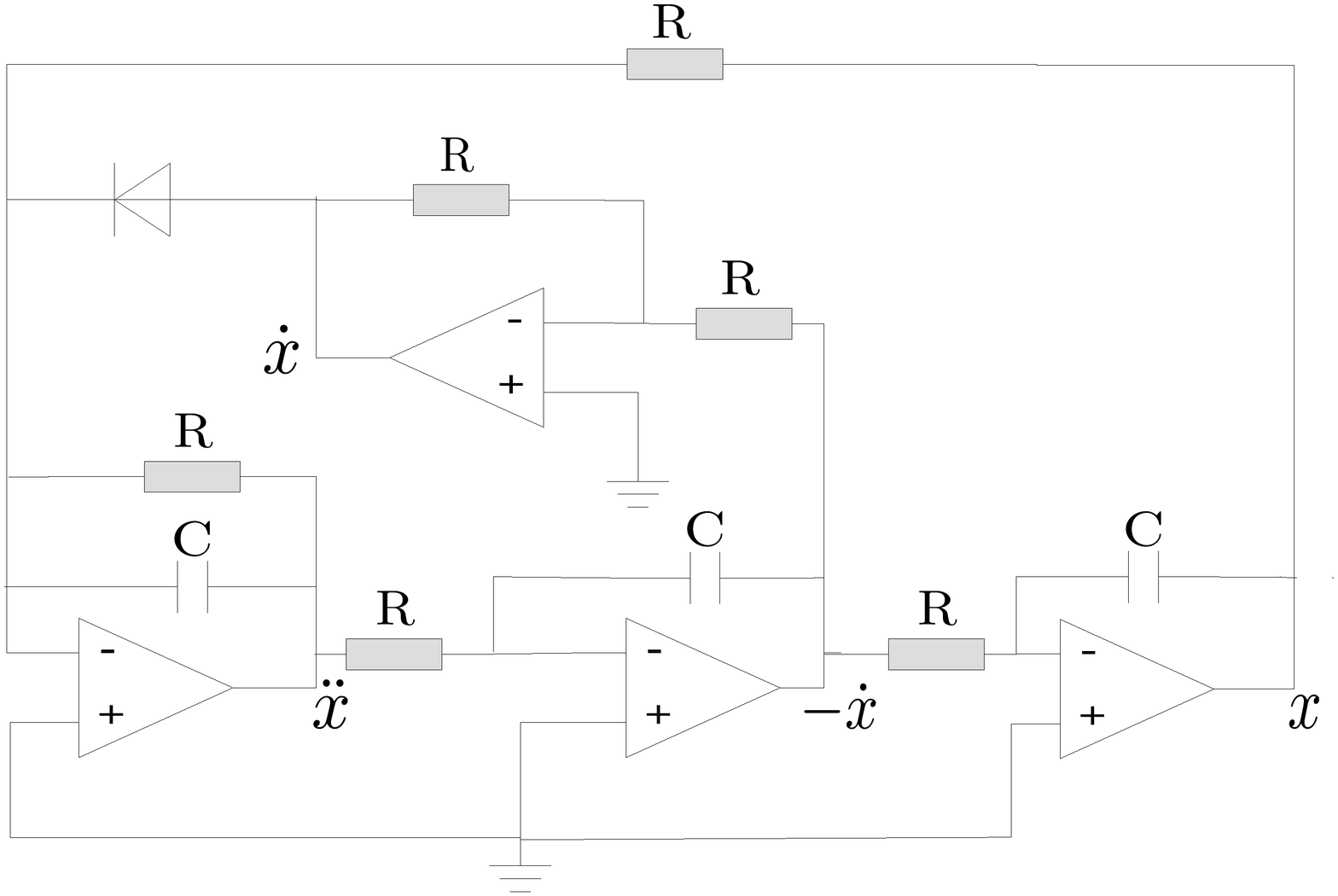}
	\caption{Circuito caótico composto por resistores de 1k$\Omega$, amplificadores operacionais, capacitores de 1$\mu$F e diodo 1N4001. Fonte: \cite{Spr2011}.}
	\label{Jerk}
\end{figure}

Desse modo, a metodologia adotada para realizar esses procedimentos é mostrada abaixo
\begin{itemize}
    \item Simulação e coleta dos dados do circuito eletrônico mostrado na Fig.(\ref{Jerk}) em diferentes computadores;
    \item Implementação prática do circuito eletrônico;
    \item Coleta dos dados da implementação via placa de aquisição de dados;
\end{itemize}

\begin{table}[!ht]
    \caption{Lista de Computadores Utilizados para Análise da Reprodubitilidade do Circuito.}
    \label{tab:comp}
    \centering
\begin{tabular}{ccc}
 Computador & Sistema & Configuração \\ 
    \hline 1  & Windows 10  & Intel Core i5 6200U  \\ 
     2 & Windows 10  &  Intel Dual core 2\\ 
     3 & Windows 8.1 & Intel Core i5-3570  \\ 
     4 & Windows 8.1  & Intel Core i5-4210U    \\\hline 
    \end{tabular}     
\end{table}

Após a realização desses procedimentos, foi feita a comparação entre resultados simulados e práticos pelo índice NRMSE.

Entretanto, esse procedimento não pode ser realizado de imediato pois como pode ser observado o cálculo do índice leva em consideração todos os dados simulados e coletados experimentalmente realizando operações matemáticas com cada um dos valores. Todos os dados, simulados e experimentais, foram armazenados em forma de vetores e para que o cálculo não apresentasse nenhum erro era necessário que esses vetores fossem de tamanhos iguais, porém, o modo inicialização e o número de pontos coletados pelo LtSpice apresentaram diferenças de um computador para o outro, resultando então vetores de tensão de tamanhos diferentes para cada computador testado. Como solução para esse problema, na rotina implementada para o cálculo do índice foi criado um novo vetor de tempo com uma quantidade de pontos estabelecida previamente, para que assim pudesse ser realizada a interpolação dos vetores originais. Ao fazer a interpolação todos os vetores de tensão coletados computacionalmente e na prática passaram a ter o mesmo número de pontos a serem analisados.

A simulação no computador que apresentou o menor NRMSE foi considerada como referência. Em seguida, partindo desse computador como referência, foi feita então a comparação com as demais simulações para verificar a reprodutibilidade. 

Com relação a propagação de erros, foram tomados os devidos cuidados seguindo as orientações contidas na norma do Instituto de Engenharia Elétrica e Eletrônica (IEEE) 754-2008 que versa sobre cálculo computacional com ponto flutuante \citep{Ove2001,IEE2008} e na norma, também do IEEE, sobre Aritmética Intervalar, a IEEE 1788-2015 \citep{Iee2015}.

\section{Resultados}

\begin{figure}[!ht]
 \hspace{-0.2cm}
	\includegraphics[width=.49\textwidth,height=.23\textheight]{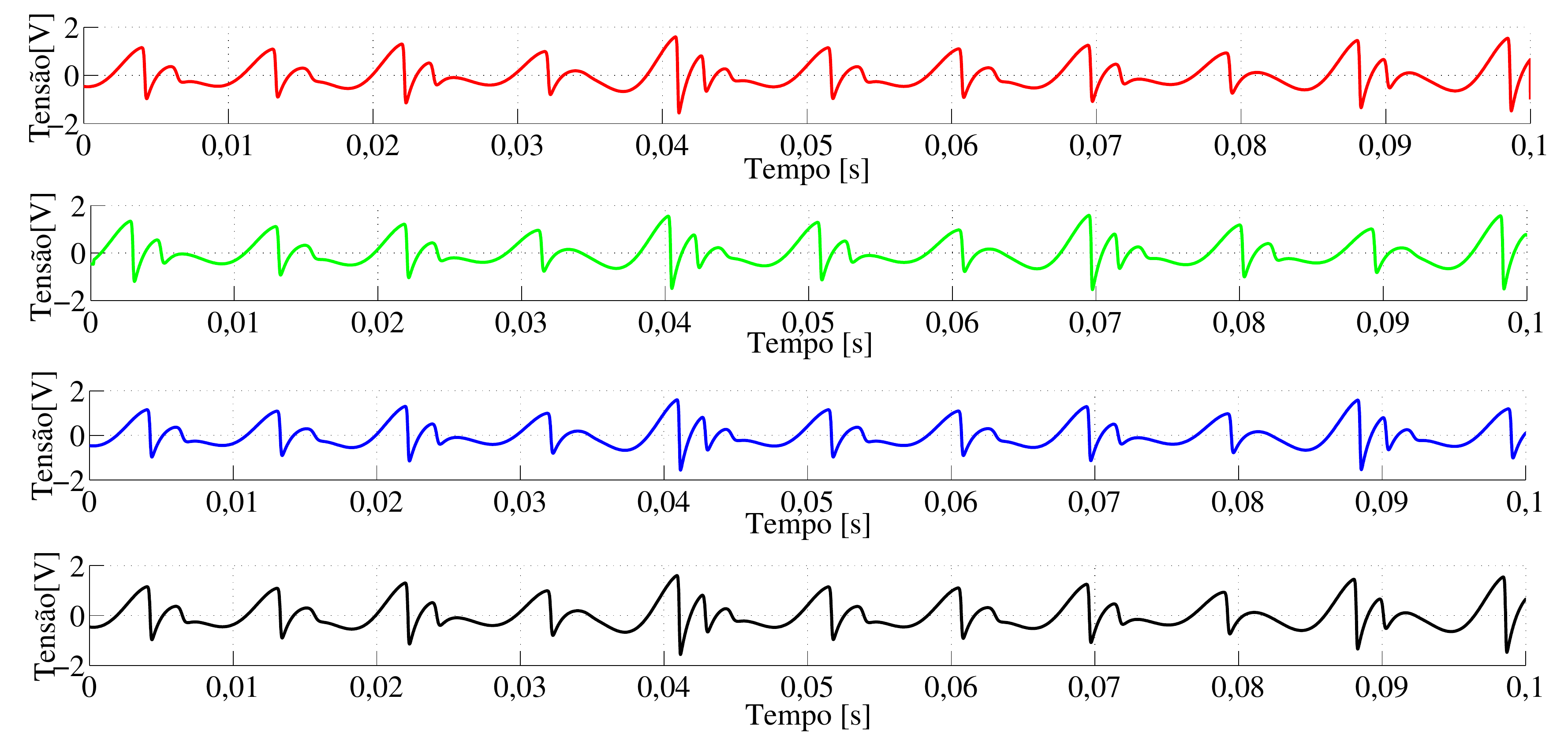}
	\caption{Tensão coletada no ponto $\ddot{x}$ de cada computador analisado. Computador 1({\color{red}\rule[0.03mm]{3mm}{0.3mm}}), Computador 2({\color{blue}\rule[0.03mm]{3mm}{0.3mm}}), Computador 3({\color{green}\rule[0.03mm]{3mm}{0.3mm}}), Computador 4(\rule[0.03mm]{3mm}{0.3mm}).  }
	\label{todos}
\end{figure}

O circuito  Jerk foi implementado no LtSpice XVII nos quatro computadores listados pela Tabela \ref{tab:comp}, utilizando a mesma versão do \textit{software} para todos os computadores. Em seguida, foram coletados os dados do circuito físico utilizando a placa de aquisição de dados. Todos os resultados foram coletados no intervalo de tempo entre 0 e 0,1 segundos. A Figura \ref{todos} apresenta a resposta para tensão coletada no ponto $\ddot{x}$ para cada computador.

Após todos os dados coletados, foi realizada a comparação entre os resultados para verificar a reprodutibilidade entre todos os resultados computacionais com o resultado da prática. 

A comparação foi realizada utilizando uma rotina no \textit{software} Matlab para o cálculo do índice NRMSE considerando todos os valores de tensão já interpolados de cada computador. Os resultados obtidos são apresentados pela Tabela \ref{nrmse1}. Analisando apenas o gráfico apresentado pela Figura (\ref{todos}) acredita-se que não há diferença nos resultados de um computador para o outro, entretanto, ao fazer a comparação dos resultados utilizando o índice NRMSE, verifica-se que entre os computadores há uma divergência de resultados, pois o valor do índice se distancia de zero.

\begin{table}[!ht]
    \caption{Resultado do índice NRMSE quando comprados dados simulados e experimentais.}
    \label{nrmse1}
    \centering
\begin{tabular}{cc}
 \textbf{Computador} & \textbf{NRMSE} \\ 
    \hline 1 &  1,4752\\
     2 &  1,5572\\
     3 &  1,4841\\
     4 &  1,4748\\
    \hline
    \end{tabular}     
\end{table}

Como dito anteriormente, o computador que apresentasse o menor valor de NRMSE seria considerado como referência. Entretanto, acredita-se que com o passar do tempo esse índice tende a aumentar. Desta forma, foi necessário um cálculo em partes para o NRMSE. Esse cálculo foi realizado da seguinte forma, cada vetor de tensão tem uma quantidade de 4700 pontos coletados. Esses pontos foram divididos em dez partes, calculando assim o índice para cada um delas, sendo o cálculo acumulativo.

Pode ser observado pela Tabela \ref{partes} com o aumento do número de pontos coletados há também o aumento do valor do índice como já esperado. Os computadores 1 e 4 foram os que apresentaram os menores valores. Entretanto, para a comparação da última parte em que são analisados todos os pontos, o computador 4 apresentou um melhor desempenho. Desta forma, ele foi escolhido como computador referência para verificar a reprodutibilidade do LtSpice.

\begin{table}[htp]
\centering
\caption{Resultado do índice NRMSE para diferentes intervalos de simulação.}
\vspace{0.5cm}
\label{partes}
\begin{tabular}{ccccc}

&\multicolumn{3}{r}{\textbf{Computadores}}\\ \hline
\multicolumn{1}{l}{\textbf{Iteração}} & \multicolumn{1}{c}{\textbf{1}} & \multicolumn{1}{c}{\textbf{2}} & \multicolumn{1}{c}{\textbf{3}} & \multicolumn{1}{c}{\textbf{4}} \\  \hline
1-470           & 0,6457    & 1,3960    & 0,6480   & 0,6457      \\ 
1-940           & 0,9843    & 1,4110    & 0,9854   & 0,9843      \\ 
1-1410          & 1,1611    & 1,4244    & 1,1616   & 1,1611      \\ 
1-1880          & 1,2396    & 1,5270    & 1,2398   & 1,2396      \\ 
1-2350          & 1,2643    & 1,5132    & 1,2649   & 1,2643      \\ 
1-2820          & 1,3197    & 1,5093    & 1,3194   & 1,3197      \\ 
1-3290          & 1,3797    & 1,5569    & 1,3835   & 1,3797      \\ 
1-3760          & 1,3993    & 1,5453    & 1,4040   & 1,3993      \\ 
1-4230          & 1,4431    & 1,5437    & 1,4567   & 1,4431      \\ 
1-4700          & 1,4752    & 1,5572    & 1,4841   & 1,4748      \\ \hline
\end{tabular}
\end{table}

\begin{figure}[htp]
	\centering
	\includegraphics[width=.49\textwidth,height=.2\textheight]{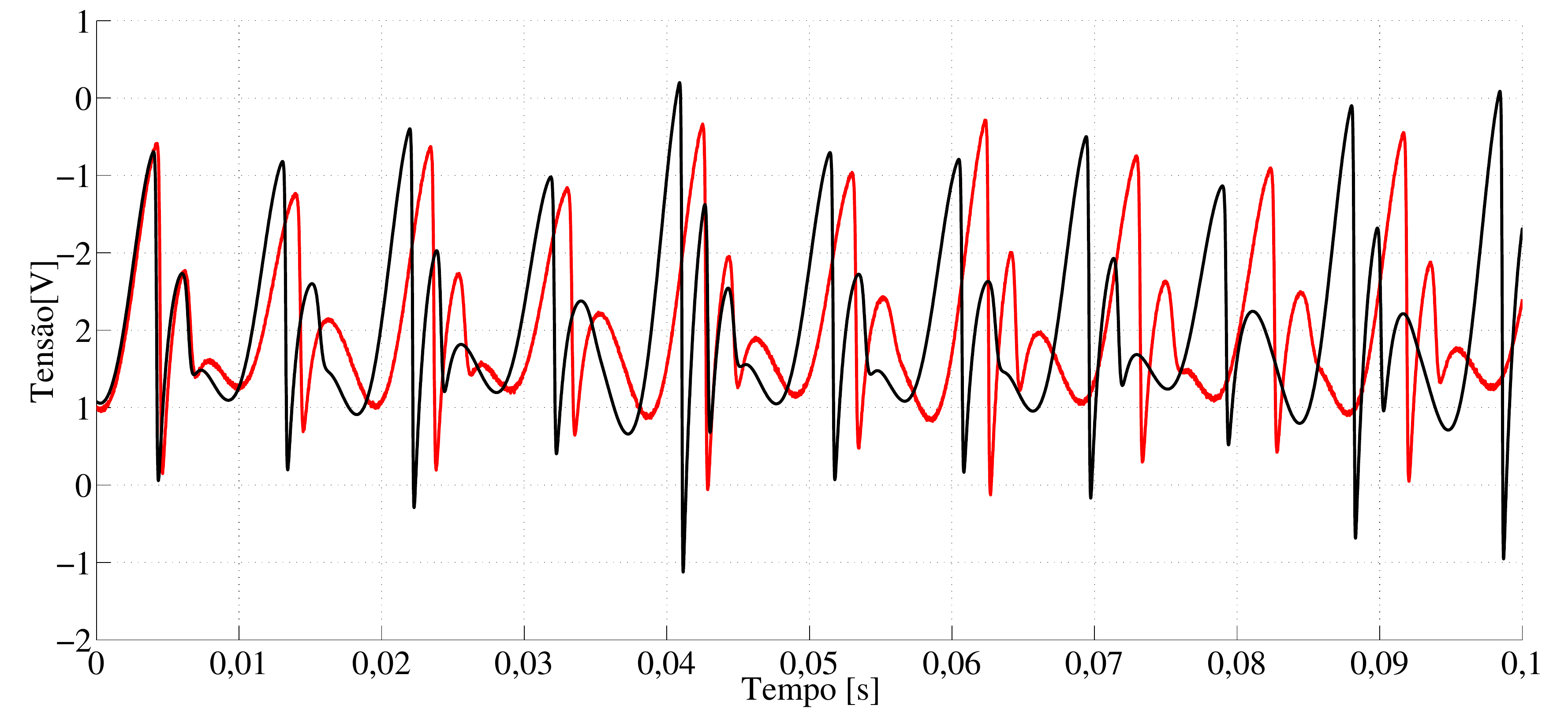}
	\caption{Tensão em $\ddot{x}$ para o computador 4, que apresentou o menor índice NRMSE, e os dados coletados experimentalmente. Computador 4({\color{red}\rule[0.03mm]{3mm}{0.3mm}}), Dados experimentais(\rule[0.03mm]{3mm}{0.3mm})}
	 \label{praticapc}
\end{figure}

Esses resultados podem ser observados na Figura \ref{praticapc}. Nos primeiros instantes os valores de tensão permanecem próximos entre o computador 4, que será adotado como referência, e os dados experimentais. E quanto maior é o tempo de simulação, mais eles vão se distanciando.

\begin{figure}[htp]
 \label{comparar}
	\centering
	\includegraphics[width=.49\textwidth,height=.2\textheight]{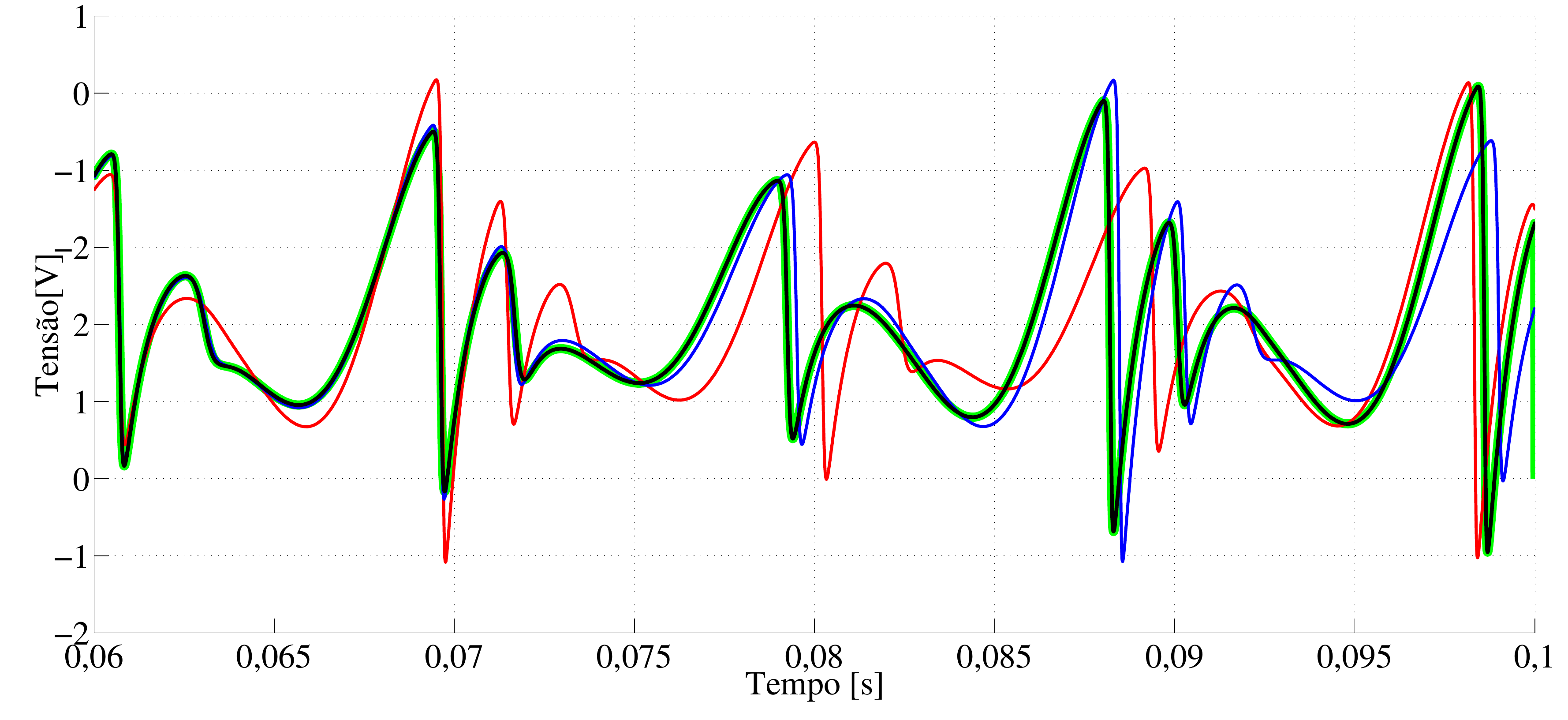}
	\caption{Tensão em $\ddot{x}$ para cada computador analisado. Computador 1({\color{green}\rule[0.03mm]{3mm}{0.3mm}}), Computador 2({\color{red}\rule[0.03mm]{3mm}{0.3mm}}), Computador 3({\color{blue}\rule[0.03mm]{3mm}{0.3mm}}), Computador 4(\rule[0.03mm]{3mm}{0.3mm}).}
\end{figure}

Após fazer o novo cálculo do NRMSE entre os computadores, utilizando o computador 4 como referência, foram obtidos os valores mostrados pela Tabela \ref{nrmse2}. Com esses resultados é possível demonstrar que a reprodutibilidade do \textit{software} é maior quando as características dos computadores se assemelham. Pode ser observado também que o computador 2 foi o que mais se distanciou do resultado prático, foi também o que apresentou menor reprodutibilidade. Entretanto, mesmo resultando em pequenos valores, foi possível notar que há diferença na resposta entre dois computadores de configurações semelhantes. Esta diferença é melhor compreendida observando a Figura \ref{comparar}.

\begin{table}[!ht]
    \caption{Resultado do índice NRMSE assumindo o computador 4 como referência.}
    \vspace{0.5cm}
    \label{nrmse2}
    \centering
\begin{tabular}{cc}
 \textbf{Computador} & \textbf{NRMSE} \\ 
    \hline 1 &   0,0454\\
     2 &   1,0128\\
     3 &   0,4414\\
    \hline
    \end{tabular}     
\end{table}

A Figura \ref{comparar} apresenta uma comparação gráfica entre os valores de tensão coletados em cada computador. É possível verificar que o Computador 1 acompanha praticamente por todo intervalo de simulação o Computador 4, como esperado, pois o valor do índice quando comparados é próximo de zero. Por outro lado, o Computador 2, logo no início da simulação, apresenta uma diferença considerável comparado aos demais computadores. 

Ao final de todas as análises foi possível observar que o mesmo problema de reprodutibilidade encontrado por \citep{Salamon}, ao tentar reproduzir o circuito de Chua utilizando o \textit{software} multsim, ocorre com o circuito Jerk, quando utilizado a mesma versão do \textit{software} LtSpice em diferentes computadores.

\section{Conclusão}

Ao final desse trabalho, foi possível observar que a princípio simulações realizadas no LtSpice podem apresentar resultados semelhantes, principalmente quando é realizada apenas a análise gráfica. Entretanto, seguindo a metodologia abordada nesse projeto, foi possível demonstrar que a simulação de um mesmo circuito sob mesmas condições iniciais e utilizando a mesma versão do \textit{software} não apresenta os mesmos resultados quando submetida à computadores diferentes. Acredita-se que isso ocorre devido ao método de simulação do \textit{software} em cada computador ocorrer de forma diferente, tanto em sua forma de inicializar a simulação, quanto no armazenamento dos resultados de cada ponto do circuito, isso pode ser mostrado com a diferença na quantidade de pontos coletados em cada simulação. Além desse fato, é necessário observar que o método utilizado para análise da reprodutibilidade não apresenta os mesmos valores para cada intervalo de tempo da simulação, sendo seu resultado nos primeiros instantes mais próximo de zero do que ao considerar todos os pontos coletados. Esse fato é importante ao realizar análises e comparações, pois, uma vez que se deseja usar os dados de um computador específico, é necessário ter a ciência de que ao trocar de computador para realizar a mesma simulação pode-se obter valores e quantidade de pontos diferentes acarretando na divergência de resultados. Logo, é necessário que em simulações de circuitos que envolvem o \textit{software} LtSpice haja um maior cuidado levando em consideração todas as observações apresentadas. 

Longe deste trabalho apresentar uma solução definitiva para o problema, acredita-se que a metodologia empregada tenha sido capaz de identificar  o problema, escolher computadores com melhor desempenho em termos de horizonte de predição. Uma outra vertente apresentada recentemente pelo IEEE que pode contribuir para este cenário trata-se do \textit{Code Ocean}. Trata-se de uma plataforma nas nuvens em que o pesquisador e o leitor do trabalho científico executa o código em um mesmo hardware e software \citep{ElHawary2018}. Certamente, traz uma perspectiva de reprodutibilidade amplamente superior ao que se tem apresentado na maioria da literatura até o momento e merece a atenção da comunidade científica. Adicionalmente, como perspectiva de trabalhos futuros, pretende-se propor um método, utilizando a simulação do expoente de Lyapunov, que indique o intervalo de tempo para que haja confiabilidade nos resultados. Isso pode ser interessante, pois mesmo no Code Ocean, differente linguagens podem oferecer resultados diferentes.

\section*{Agradecimentos}
Os autores agradecem ao apoio financeiro da Fapemig, CNPq, Capes e à Universidade Federal de São João del-Rei.

\bibliography{jerk}             % bib file to produce the bibliography

\end{document}